\documentclass[english,sort&compress, reprint, superscriptaddress]{revtex4-1}

\usepackage[T1]{fontenc}
\usepackage[utf8]{inputenc}
\usepackage{textcomp}
\usepackage{graphicx}
\usepackage{subscript}

\makeatletter
\usepackage{hyperref}
\usepackage{amssymb,amsmath}

\makeatother

\usepackage{babel}
\begin{document}
\title{Unidirectional Nano-Modulated Binding and Electron Scattering in Epitaxial
Borophene}
\author{Sherif Kamal}
\affiliation{Centre for Advanced Laser Techniques, Institute of Physics, 10000
Zagreb, Croatia}
\author{Insung Seo}
\affiliation{Department of Materials Science and Engineering, Tokyo Institute of
Technology, Yokohama 226-8502, Japan}
\author{Pantelis Bampoulis}
\affiliation{Physics of Interfaces and Nanomaterials, MESA+ Institute, University
of Twente, 7522 NB, Enschede, The Netherlands}
\affiliation{Institute of Physics II, University of Cologne, 50937 Cologne, Germany}
\author{Matteo Jugovac}
\affiliation{Elettra - Sincrotrone Trieste S.C.p.A., S.S. 14 km 163.5, 34149 Trieste,
Italy}
\author{Carlo Alberto Brondin}
\affiliation{Department of Molecular Sciences and Nanosystems, Ca’ Foscari University
of Venice, 30172 Venice, Italy}
\author{Tevfik Onur Mente{\c{s}}}
\affiliation{Elettra - Sincrotrone Trieste S.C.p.A., S.S. 14 km 163.5, 34149 Trieste,
Italy}
\author{Iva Šarić Janković}
\affiliation{Faculty of Physics and Center for Micro- and Nanosciences and Technologies,
University of Rijeka, 51000 Rijeka, Croatia}
\author{Andrey V. Matetskiy}
\affiliation{Istituto di Struttura della Materia-CNR (ISM-CNR), S.S. 14 km 163.5,
34149, Trieste, Italy}
\author{Paolo Moras}
\affiliation{Istituto di Struttura della Materia-CNR (ISM-CNR), S.S. 14 km 163.5,
34149, Trieste, Italy}
\author{Polina M. Sheverdyaeva}
\affiliation{Istituto di Struttura della Materia-CNR (ISM-CNR), S.S. 14 km 163.5,
34149, Trieste, Italy}
\author{Thomas Michely}
\affiliation{Institute of Physics II, University of Cologne, 50937 Cologne, Germany}
\author{Andrea Locatelli}
\affiliation{Elettra - Sincrotrone Trieste S.C.p.A., S.S. 14 km 163.5, 34149 Trieste,
Italy}
\author{Yoshihiro Gohda}
\affiliation{Department of Materials Science and Engineering, Tokyo Institute of
Technology, Yokohama 226-8502, Japan}
\author{Marko Kralj}
\affiliation{Centre for Advanced Laser Techniques, Institute of Physics, 10000
Zagreb, Croatia}
\author{Marin Petrović}
\email{mpetrovic@ifs.hr}

\affiliation{Centre for Advanced Laser Techniques, Institute of Physics, 10000
Zagreb, Croatia}
\begin{abstract}
A complex interplay between the crystal structure and electron behavior
within borophene renders this material an intriguing 2D system with
many of its electronic properties still undiscovered. Experimental
insight into those properties is additionally hampered by the limited
capabilities of the established synthesis methods, which in turn inhibits
the realization of potential borophene applications. In this multi-method
study, photoemission spectroscopies and scanning probe techniques
complemented by theoretical calculations have been used to investigate
the electronic characteristics of a high-coverage, single-layer borophene
on Ir(111) substrate. Our results show that the binding of borophene
to Ir(111) exhibits pronounced one-dimensional modulation and transforms
borophene into a nano-grating. The scattering of photoelectrons from
this structural grating gives rise to the replication of electronic
bands. In addition, the binding modulation is reflected in the chemical
reactivity of borophene and gives rise to its inhomogeneous aging
effect. Such aging is easily reset by dissolving boron atoms in iridium
at high temperature followed by their reassembly into a fresh atomically-thin
borophene mesh. Besides proving electron-grating capabilities of the
boron monolayer, our data provides a comprehensive insight into the
electronic properties of epitaxial borophene which is vital for further
examination of other boron systems of reduced dimensionality.
\end{abstract}
\maketitle

\section{Introduction}

Synthesis of various borophene (Bo) polymorphs on a range of different
substrates \citep{Li2023} opened new pathways for investigation of
these atomically-thin boron systems. In order to attain precise control
over the synthesis, growth of Bo is typically carried out in ultra-high
vacuum (UHV) conditions on metallic single-crystals. In such way,
epitaxial Bo has been realized on Ag, Cu, Ir, Au, Al and Ru substrates
\citep{Feng2016c,Wu2019,Vinogradov2019,Kiraly2019,Li2018,Sutter2021}.
Epitaxial growth holds potential for high sample quality – in terms
of size, number of defects and structural homogeneity – which is often
an imperative for investigation of intrinsic properties of Bo. Also,
high material quality is required for various technological applications
of Bo, such as metal-ion batteries \citep{Zhang2016,Yu2022}, supercapacitors
\citep{Li2018d}, hydrogen storage \citep{Er2009}, gas sensors \citep{Huang2018}
and freshwater production \citep{Guan2023}.

However, binding to the metal substrate is reflected in the structural
and physical properties of epitaxial Bo. Interaction of Bo with the
underlying metal not only defines the most stable polymorph \citep{Zhang2015e},
but can also induce new morphology in the form of stripes, chains
and zig-zag arrays which are identified as the moiré patterns in some
cases \citep{Feng2016c,Wu2019,Liu2019d,Kiraly2019,Wang2020}. All
these nanoscopic structures are in turn linked to the electronic and
vibrational properties of Bo, and are expected to perturb intrinsic
features of Bo such as high electrical and thermal conductivity \citep{Adamska2018,Zhou2017}.
Signatures of nano-modulation in other epitaxial two-dimensional (2D)
systems can readily be found in the spectroscopic data, namely the
valence band and the core-level electronic structure. Examples of
this are Dirac cone replication of graphene on Ir(111) \citep{Pletikosic2009}
or distinct peaks in the N $1s$ spectra of hexagonal boron nitride
(hBN) on transition metals \citep{Preobrajenski2007}. However, experimental
detection of similar signatures in Bo samples is only scarcely addressed
in the literature.

In particular, studies employing angle-resolved photoemission spectroscopy
(ARPES), the best tool for studying the valence band structure of
materials, are largely lacking. ARPES has been employed in only a
few Bo investigations due to the experimental challenges arising from
the sample inhomogeneity (e.g. presence of several polymorphs, small
domains, defects) and significant chemical reactivity of Bo. Up to
now, only $\beta_{12}$ and $\chi_{3}$ polymorphs on Ag(111) have
been examined with ARPES by Feng and co-workers, where metallicity
and the existence of the Dirac cones have been suggested \citep{Feng2016,Feng2017a,Feng2018}.
This limited set of data, especially considering a large number of
other realizable polymorphs, leaves many open questions related to
the repercussions of Bo nano-modulation on its electronic properties
and performance in electronic devices.

Here we report a combined experimental and computational study of
Bo on Ir(111) (Bo/Ir) with a focus on the electronic properties governed
by the nanoscopic modulation of the boron sheet. X-ray photoelectron
spectroscopy (XPS) and scanning tunneling spectroscopy (STS) data
show clear evidence of inhomogeneous binding of Bo to Ir, which results
in a stripe-like structure of Bo. This structure in turn effectively
produces a one-dimensional (1D) grating, giving rise to the umklapp
scattering of photoelectrons which are detected in ARPES experiments.
Density functional theory (DFT) calculations support our experimental
findings and give an additional insight into the pristine electronic
structure of Bo, void of nanoscopic modulation imprinted by the presence
of the Ir substrate.

\section{Results}

\subsection{Evolution of the Ir $4f$ and B $1s$ core-levels}

\begin{figure*}
\begin{centering}
\includegraphics{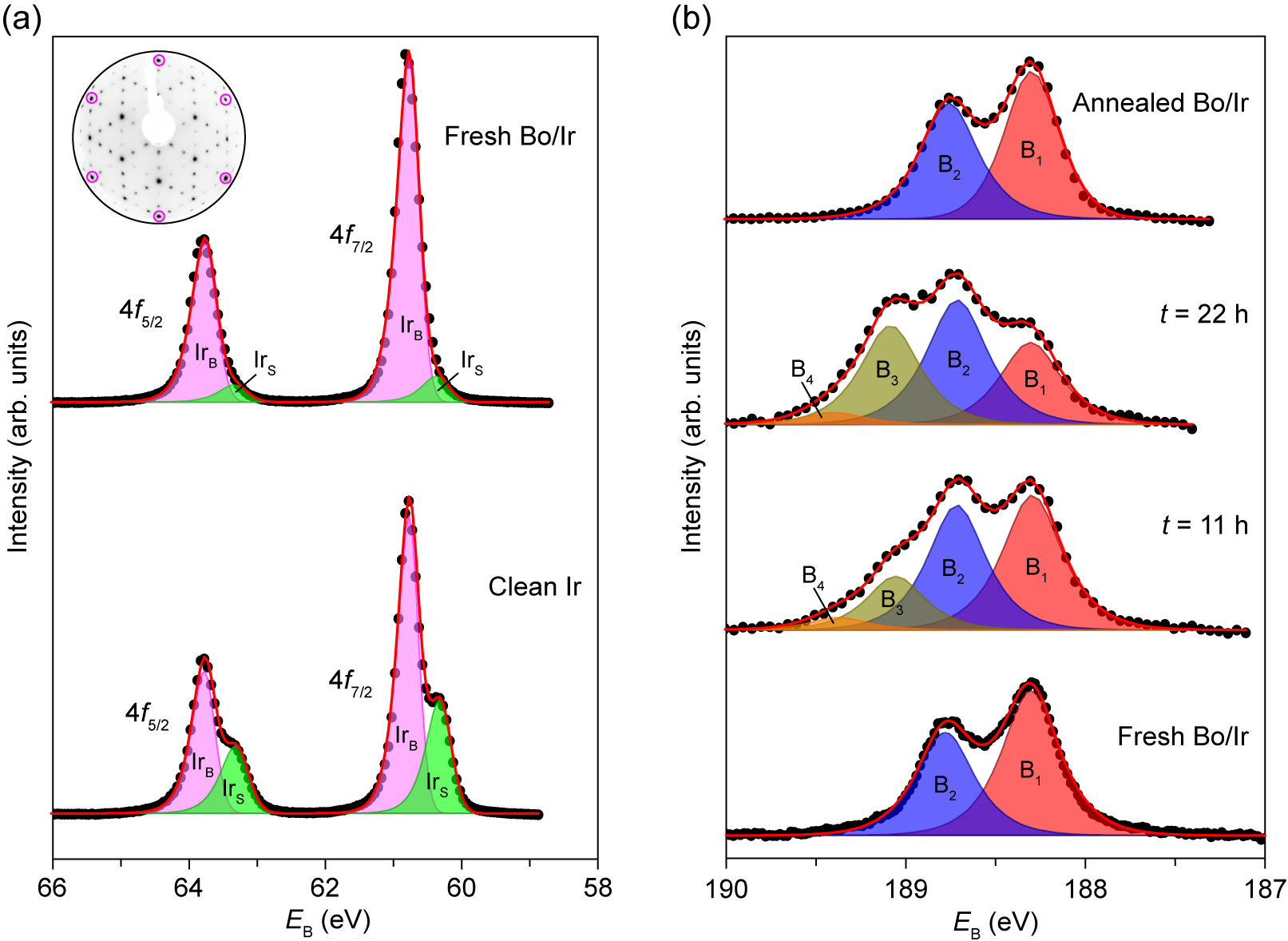}
\par\end{centering}
\caption{\label{figXPS}Core-level spectra before and after Bo synthesis on
Ir(111). (a) Ir $4f$ spectra measured for clean Ir (bottom) and Bo/Ir
(top) samples, with the bulk (Ir\protect\textsubscript{B}) and surface
components (Ir\protect\textsubscript{S}) indicated. The inset shows
LEED pattern of Bo/Ir recorded at 83 eV of electron energy, with the
Ir diffraction spots encircled in magenta. (b) B $1s$ spectra of
the Bo/Ir sample recorded immediately after Bo synthesis (bottom),
11 and 22 hours after the synthesis (middle) and after sample flash-annealing
to 1100 °C (top). Different components of the B $1s$ level (B\protect\textsubscript{1...4})
are indicated. All spectra are plotted with a Shirley background subtracted
and normalized to the total intensity under the fitted curve (red
line). All spectra were recorded with $p-$polarized light at $h\nu=300$
eV.}
\end{figure*}

Prior to Bo synthesis, a reference Ir $4f$ core-level photoemission
spectrum has been recorded as shown in Fig. \ref{figXPS}(a) bottom.
Surface (Ir\textsubscript{S}) and bulk (Ir\textsubscript{B}) components
of the Ir $4f$ peaks are discerned with binding energies ($E_{\mathrm{B}}$)
of 60.33 and 60.80 eV (Ir $4f_{7/2}$), and 63.33 and 63.80 eV (Ir
$4f_{5/2}$), respectively, in agreement with the literature data
\citep{Varykhalov2012a}. The intensity ratio $I_{\mathrm{IrS}}/I_{\mathrm{IrB}}$
of surface to bulk components, at the utilized photon energy of 300
eV, is 0.38 for Ir $4f_{7/2}$ and 0.42 for Ir $4f_{5/2}$.

Bo synthesis consisted of B uptake in the Ir bulk during exposure
of hot Ir crystal to the borazine vapors, followed by sample cooling
which resulted in segregation of B atoms to the surface and their
self-assembly into a Bo mesh (see Methods for further details) \citep{Omambac2021}.
Successful Bo synthesis was firstly confirmed by LEED, where a characteristic
$\left(6\times2\right)$ pattern in three rotational domains was found
{[}see Fig. \ref{figXPS}(a) inset{]}, designating the presence of
the $\chi_{6}$ polymorph on the sample surface. Comparable intensities
of Ir and Bo diffraction spots are an indication of relatively high
Bo coverage \citep{Radatovic2022}. The respective Ir $4f$ core level
spectra of Bo/Ir is shown in Fig. \ref{figXPS}(a) top. In order to
get a reliable fit, both Ir\textsubscript{S} and Ir\textsubscript{B}
components must be included in the fitting procedure. The binding
energies of all components remain the same as before Bo synthesis,
but the $I_{\mathrm{IrS}}/I_{\mathrm{IrB}}$ ratios reduce significantly
(by a factor $\sim4$) for both the Ir $4f_{7/2}$ and the Ir $4f_{5/2}$
components.

Apparently, the presence of Bo on top of Ir surface suppresses the
surface component of the Ir $4f$ level. This indicates significant
interaction between the Bo $\chi_{6}$ polymorph and Ir(111), since
a similar reduction of the Ir\textsubscript{S} peak is detected after
adsorption of strongly bound species to the Ir surface \citep{Bianchi2009,Baronio2022}.
In contrast, for weakly interacting epitaxial layers, such as graphene
or hBN on Ir(111) \citep{Varykhalov2012a,FarwickZumHagen2016}, the
surface component is almost unaltered in the presence of the 2D overlayer.
Possible origin of the relatively small surface component which persists
in our sample could be in the fraction of the Ir surface which is
left uncovered after the Bo synthesis, considering recent micro-spot
XPS examination of Bo/Ir where only bulk Ir $4f$ peaks have been
detected \citep{Jugovac2023}.

B $1s$ spectrum of the Bo/Ir(111) sample recorded immediately after
Bo synthesis is shown in Fig. \ref{figXPS}(b) bottom. Two distinct
components are visible, labeled B\textsubscript{1} and B\textsubscript{2},
located at $E_{\mathrm{B}}$ of 188.30 and 188.78 eV. The intensity
ratio of the two components is $I_{\mathrm{B1}}/I_{\mathrm{B2}}=1.4$.
For the freshly synthesized Bo samples, no other B-related peaks were
observed. Also, no signatures were found of additional B peaks at
higher binding energies, characteristic for hBN layers \citep{Orlando2011}.

The B $1s$ energy region has been monitored via XPS over an extended
period of time during which the sample was left in the UHV chamber
at a base pressure in the 10\textsuperscript{-10} mbar range. As
shown in Fig. \ref{figXPS}(b) for samples 11 and 22 hours after the
synthesis, we observed modifications of the initial B\textsubscript{1}
and B\textsubscript{2} peaks as well as the appearance of additional
peaks. The B\textsubscript{1} peak remains at the same energy but
loses intensity as the time progresses. The B\textsubscript{2} peak
shows a different behavior – it maintains an almost constant intensity
but slightly shifts, by $74$ meV, to lower binding energies. After
22 hours, the ratio $I_{\mathrm{B1}}/I_{\mathrm{B2}}$ decreases to
0.65. The reduction of the B\textsubscript{1} peak intensity is accompanied
by the emergence of B\textsubscript{3} and B\textsubscript{4} components
at 189.07 eV and 189.37 eV, respectively. These modifications of the
B $1s$ peak are attributed to binding of residual C- and/or O-containing
molecules from the UHV environment to boron atoms \citep{Moddeman1989,Ong2004,Cattelan2013}.
Based on the B 1$s$ peak intensities, we conclude that after 22 hours
in UHV these molecules bind to approximately $1/3$ of boron atoms
within Bo. Significant chemical reactivity of Bo sheets contrasts
chemical inertness of boron-containing hBN which shows no contamination
signatures in XPS data even after prolonged exposure to air \citep{Kidambi2014}.
It was demonstrated that exactly such inertness of hBN, as well as
that of graphene, can be utilized for protecting Bo from contamination
by synthesizing hBN/Bo and graphene/Bo vertical heterostructures \citep{Cuxart2021,Jugovac2023}.

In the follow-up control experiment, the sample was flashed to 1100
°C and immediately cooled down, in order to dissolve boron in Ir,
desorb any contaminants from the sample surface and produce a fresh
Bo layer via B segregation to the Ir surface (see Supporting Figure
S1). Such procedure is a straightforward route for obtaining pristine
samples, relying on the temperature-dependent solubility of B in Ir
that was already exploited in previous studies \citep{Omambac2021,Omambac2023}.
Indeed, the XPS spectrum of the annealed sample shown in Fig. \ref{figXPS}(b)
top looks essentially the same as the spectrum of the initial, pristine
Bo/Ir sample, exhibiting only B\textsubscript{1} and B\textsubscript{2}
components with restored positions and intensities. In addition, LEED
data exhibiting a sharp $(6\times2)$ pattern in three domains (not
shown) confirmed the presence of a clean Bo layer.

\subsection{Valence band structure}

\begin{figure*}
\begin{centering}
\includegraphics{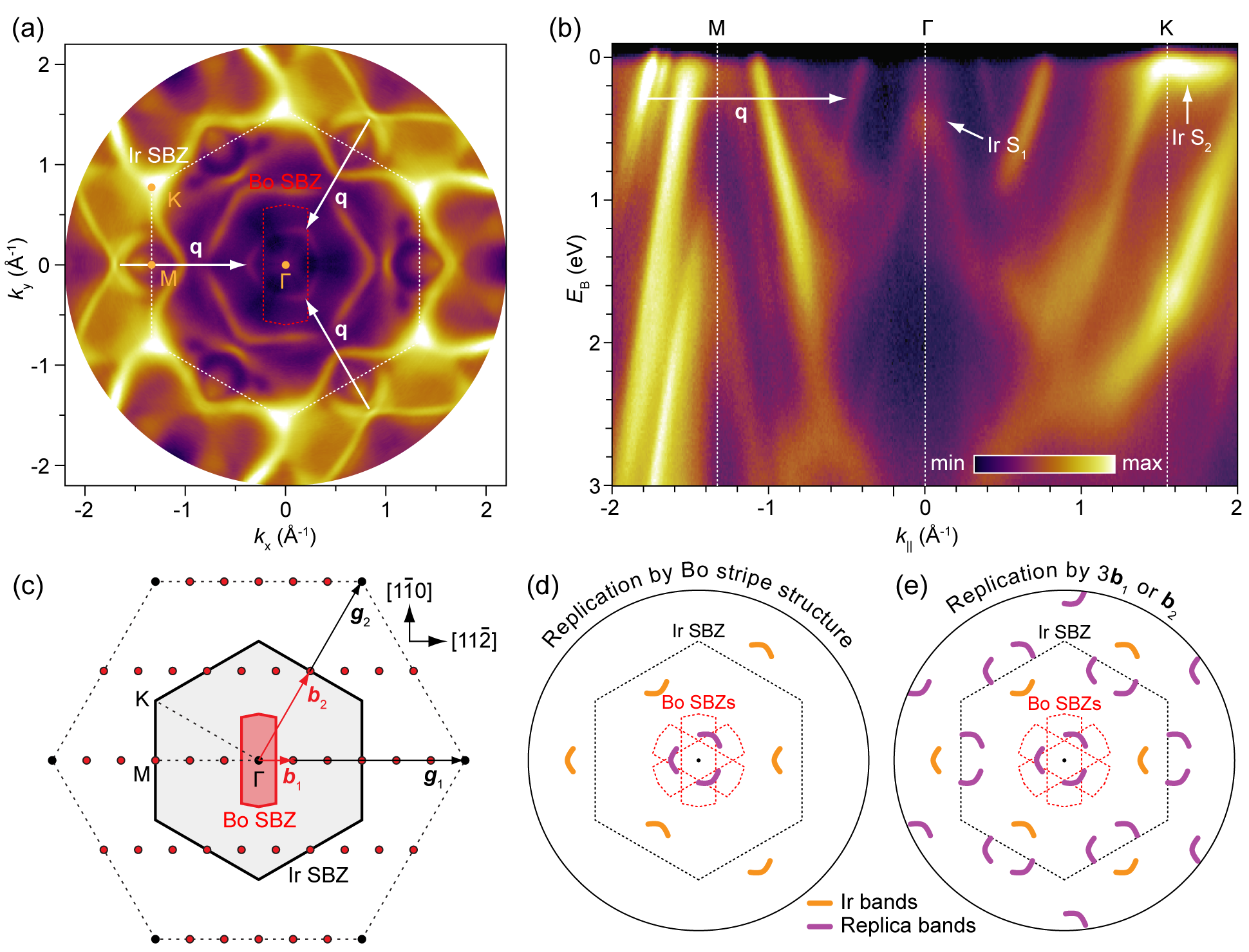}
\par\end{centering}
\caption{\label{figCARPES}ARPES data of epitaxial Bo on Ir(111). (a) Fermi
surface of Bo/Ir, with the first SBZ of Ir(111) (white regular hexagon)
and Bo (red elongated hexagon) indicated. Arrows point to Bo-induced
replica bands and designate the replication wavevectors $\mathbf{q}$.
(b) Map along the M-$\Gamma$-K direction with the replication wavevector
$\mathbf{q}$ and Ir surface states (Ir S\protect\textsubscript{1}
and Ir S\protect\textsubscript{2}) indicated. (c) Schematics of the
$k$-space of Bo/Ir. \textbf{$\mathbf{g_{\mathrm{1,2}}}$} and \textbf{$\mathbf{b_{\mathrm{1,2}}}$}
are vectors of the reciprocal unit cells of Ir and Bo, respectively.
Gray and red hexagons indicate the SBZs of Ir and Bo, respectively.
Reciprocal lattice points of Ir and Bo are marked by dots. In panels
(a) and (c), the other two 120°-rotated SBZs of Bo are not shown for
clarity.(d-e) Schematic representations of the Fermi surfaces, containing
only the electronic bands involved in the replication, in the case
of band replication by the Bo stripes (stripe-dependent) and by the
reciprocal lattice vectors $3\mathbf{b}_{1}$ or $\mathbf{b}_{2}$
(stripe-independent), respectively. In panels (d-e), all three SBZs
of Bo are indicated by red hexagons. The ARPES spectra were recorded
with $p-$polarized light at $h\nu=100$ eV.}
\end{figure*}

In order to identify any new valence band features related to Bo,
we first examined the Fermi surface of a freshly prepared sample via
spatially-averaging ARPES where the spot size on the sample surface
had a diameter of $\sim0.5$ mm. Besides the rich band structure characteristic
of the Ir(111) surface, new prominent arc-like features are found
close to the surface Brillouin zone (SBZ) center, marked by white
arrows (i.e. wavevectors $\mathbf{q}$, to be elaborated below) in
Fig. \ref{figCARPES}(a). A detailed comparison of clean Ir and Bo/Ir
ARPES data is given in the Supporting Figure S2. These arcs exhibit
anisotropic intensity and an overall 3-fold symmetric arrangement
in the $k$-space. The energy dispersion of the arcs in $\Gamma$-M
and $\Gamma$-K directions is disclosed in Fig. \ref{figCARPES}(b)
where a linear dispersion is identified extending down to $E_{\mathrm{B}}\sim1$
eV.

The Bo-induced bands are situated outside of the first SBZ of Bo and
the shape of the arcs does not reflect the expected shape imposed
by the symmetry of SBZs of Bo, which indicates that they are actually
not electronic bands of the Bo itself. By taking into account the
corrugated morphology of Bo on Ir(111) \citep{Vinogradov2019,Omambac2021},
it is plausible that the observed bands are replicated (i.e., umklapp)
bands of Ir. A similar band replication has been observed before in
other 2D material systems, e.g. for silicene on Ag(111) \citep{Sheverdyaeva2017}
and TaS on Au(111) \citep{Dombrowski2021}. Indeed, a careful inspection
of the data from Fig. \ref{figCARPES}(a) and (b) allows identification
of the Ir bands with the same Fermi contours and band dispersions
as the Bo-induced bands. The tendency of these particular Ir bands
to undergo replication might be correlated with their significant
surface localization \citep{Bornemann2012,DalCorso2015}. The corresponding
replication wavevectors $\mathbf{q}$ are noted in Fig. \ref{figCARPES}(a)
and (b) where they point from the original bands to the replica bands.
For the clarity of data representation, we choose bands outside of
the first SBZ of Ir as bands being replicated, whereas the same bands
can be found inside of the first SBZ due to $k$-space periodicity.
Schematic of the electronic bands involved in the replication is shown
in Fig. \ref{figCARPES}(d). The wavevectors $\mathbf{q}$ are parallel
to the $\Gamma$-M directions (i.e. $\left\langle 11\overline{2}\right\rangle $
directions) and have a magnitude of $1.34\pm0.01$ Å\textsuperscript{-1}.
As will be shown later, $\mathbf{q}$ is closely related to the specific
structure of Bo on Ir. Therefore, we identify the bands marked by
white wavevectors $\mathbf{q}$ in Fig. \ref{figCARPES}(a) and (b)
as replicas of the Ir bands.

The replica bands persist in UHV conditions, essentially unchanged
in position and intensity for up to 10 hours after the sample synthesis
when significant chemical modifications already occur in Bo {[}cf.
Fig. \ref{figXPS}(b){]}. This additionally supports the replication
scenario, since chemical modification of Bo is expected to also leave
an imprint on its valence band structure. Further details about band
replication process will be given in the Discussion. 

As already mentioned, the surface states of Ir are visible in the
ARPES data in Fig. \ref{figCARPES}(b) (see arrows labeled Ir S\textsubscript{1}
and Ir S\textsubscript{2}) \citep{Pletikosic2010}. The state at
the $\Gamma$ point, having an offset from $E_{\mathrm{F}}$, is more
suitable for further quantitative analysis. We find that the energy
of the surface state remains the same after Bo formation, but its
intensity is reduced and the corresponding linewidth becomes larger,
indicating reduction of the coherency of the surface state {[}see
Supporting Figure S3(a){]}. Even though contributions of the uncovered
Ir surface to these effects cannot be excluded, it is also plausible
that Bo has a specific impact on the surface state, as will be elaborated
below.

\begin{figure*}
\begin{centering}
\includegraphics{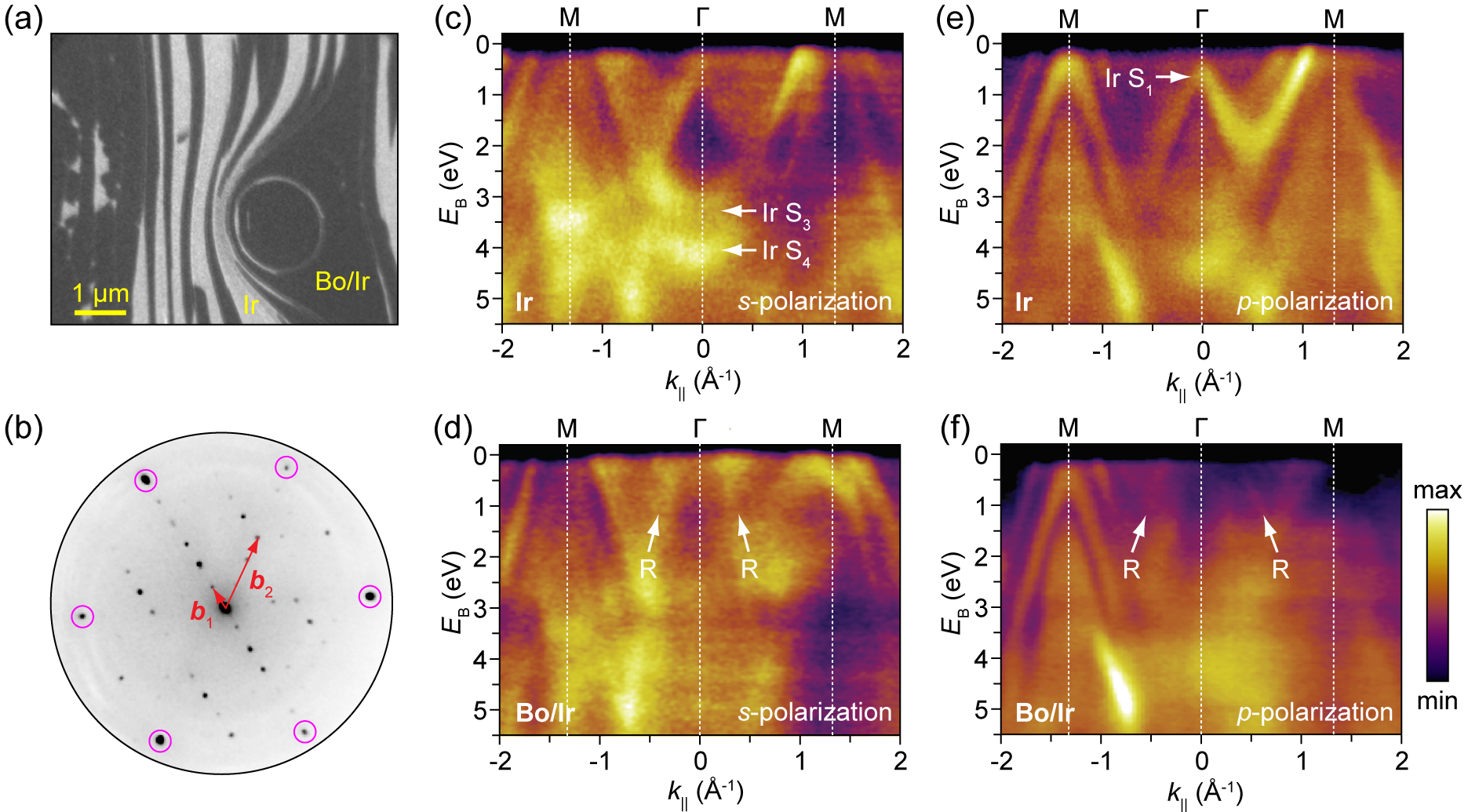}
\par\end{centering}
\caption{\label{figUARPES}$\mu-$ARPES characterization of submonolayer Bo
on Ir(111). (a) Representative LEEM image of the examined area, with
uncovered Ir (high intensity) and Bo/Ir (low intensity) regions indicated.
$V_{\mathrm{st}}=17$ eV. (b) $\mu-$LEED pattern of Bo/Ir, exhibiting
a $\left(6\times2\right)$ superstructure. The reciprocal unit cell
vectors of Bo, $\mathbf{b}_{\mathrm{1}}$ and $\mathbf{b_{\mathrm{2}}}$,
are marked by red arrows. Ir diffraction spots are encircled in magenta.
$V_{\mathrm{st}}=41$ eV. (c-f) $\mu-$ARPES spectra of uncovered
Ir and Bo/Ir regions of the sample, recorded with $p-$ and $s-$polarized
light at $h\nu=40$ eV along the M-$\Gamma$-M direction. Arrows indicate
Ir states (Ir S\protect\textsubscript{1}, Ir S\protect\textsubscript{3}
and Ir S\protect\textsubscript{4}) and Bo-induced replica bands (R).}
\end{figure*}

Spatially-resolved insight into the valence band structure of Bo on
Ir has been gained from $\mu-$ARPES measurements which complement
conventional (i.e. spatially-averaging) ARPES data. Fig. \ref{figUARPES}(a)
shows a representative area of the sample which has been deliberately
prepared with submonolayer Bo coverage for the purpose of these measurements.
Terrace-filling patterns, where a single terrace is most often filled
by a single Bo domain, are clearly resolved which is characteristic
for segregation-assisted growth of Bo on Ir(111) \citep{Omambac2021}.
Typical width of the terraces and thus individual Bo domains was of
the order of 1 $\mu$m or less, which prevented probing of a single
Bo domain with $\mu-$ARPES (having probing area diameter of $\sim2$
$\mu$m). The presence of Bo on the surface is easily identified in
$\mu-$LEED by a characteristic $\left(6\times2\right)$ pattern {[}see
Fig. \ref{figUARPES}(b){]}. Uncovered Ir and Bo/Ir regions were targeted
in $\mu-$ARPES measurements.

The corresponding valence band structures are displayed in Fig. \ref{figUARPES}(c-f).
By utilizing both $s-$ and $p-$polarized light, the Ir replica bands
can be identified in the Bo/Ir spectra which are absent in the case
of uncovered Ir surface. The iridium surface state at the $\Gamma$
point Ir S\textsubscript{1} is significantly suppressed under the
Bo layer and is almost invisible at the employed photon energy and
spectral resolution. Similar observation can be made for two other
states Ir S\textsubscript{3} and Ir S\textsubscript{4} which exhibit
significant localization in the top surface layers of iridium \citep{DalCorso2015}.
In the spectrum in Fig. \ref{figUARPES}(f), where $p-$polarization
of light has been used, a subtle increase of the photoemission intensity
is visible in the vicinity of the $\Gamma$ point, in the range $E_{\mathrm{B}}\sim3-5$
eV {[}see also Supporting Figure S3(b){]}. We speculate that this
signal can be attributed to traces of the Bo band identified recently
for graphene-capped Bo sample \citep{Jugovac2023}, which must have
$p-$orbital character since it is invisible while utilizing $s-$polarized
photons. 

$\mu-$ARPES measurements also allow the detection of secondary electron
cut-off and the determination of the work function of the sample.
It is found that the work function of Bo/Ir is $\Phi_{\mathrm{Bo/Ir}}=5.30$
eV, which is less than the work function of clean Ir(111) surface
$\Phi_{\mathrm{Ir(111)}}=5.78$ eV \citep{Derry2015}. Such work function
reduction can be assigned to hole doping of Bo, i.e. net electronic
charge transfer from Bo to Ir, similar as for graphene on metals \citep{Khomyakov2009}.
Even though marginal hole doping of Bo on Ir(111) has been theoretically
predicted \citep{Vinogradov2019}, we note that our work function
data has been recorded 4 hours after the Bo synthesis, when already
different adsorbates from the UHV could have affected the work function
value of Bo/Ir. For comparison, electron doping of Bo (accompanied
by an increase of the work function) has been measured for $\beta_{12}$
and $\chi_{3}$ polymorphs on Ag(111) \citep{Liu2021}, which demonstrates
the influence of the substrate material (and the respective binding)
on the charge rearrangement in epitaxial Bo layers.

\subsection{Microscopy and spectroscopy at the nanoscale}

\begin{figure*}
\begin{centering}
\includegraphics{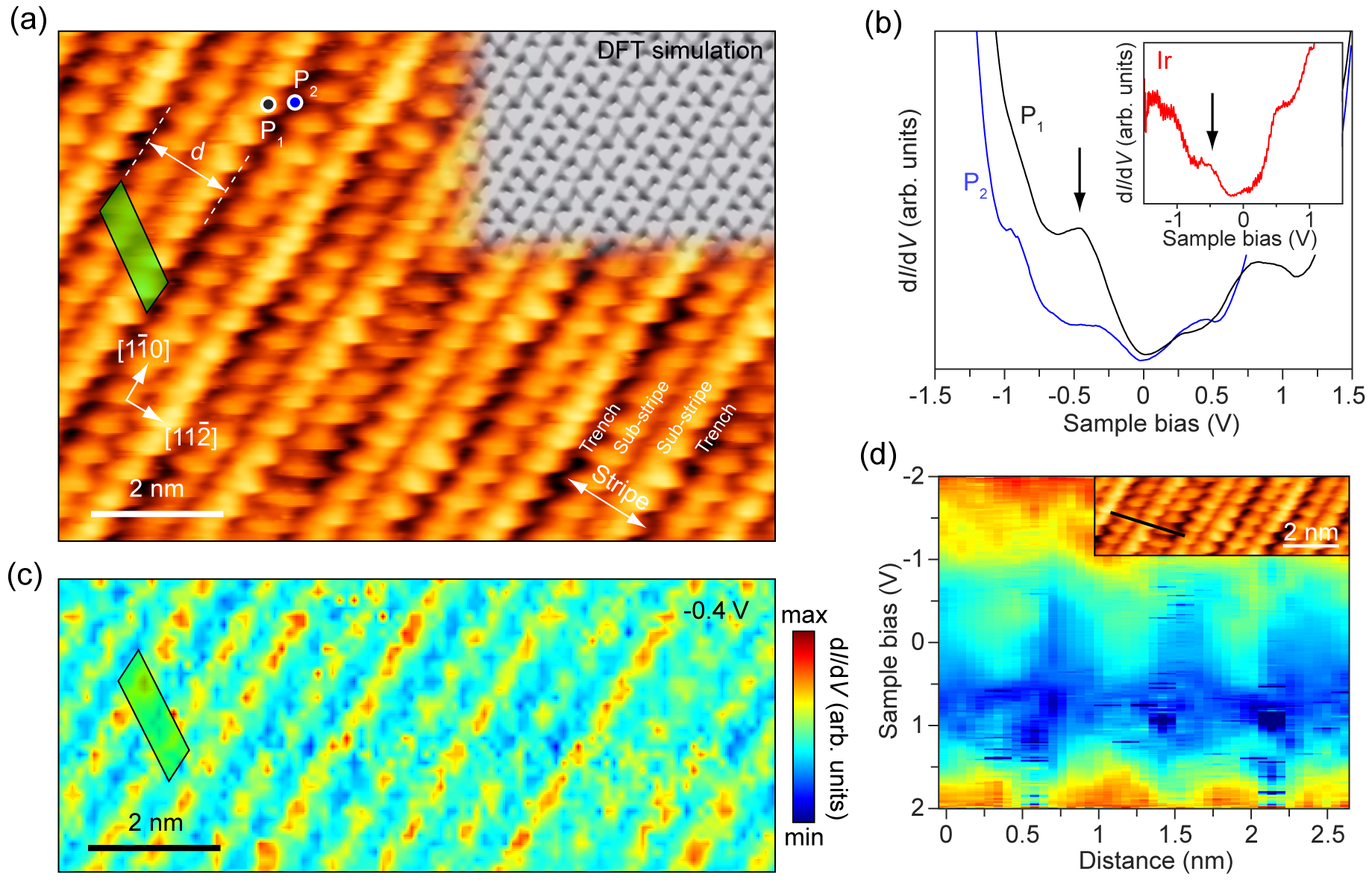}
\par\end{centering}
\caption{\label{figSTM-STS}Scanning tunneling microscopy and spectroscopy
of Bo on Ir(111). (a) High-resolution STM topograph, disclosing a
stripe-like structure of the Bo layer, with $d$ designating stripe
width. Set points: -0.5 V, 0.5 nA. Top right corner: DFT-simulated
STM image of a freestanding $\chi_{6}$ Bo polymorph. (b) $\mathrm{d}I/\mathrm{d}V$
spectra recorded at the positions P\protect\textsubscript{1} and
P\protect\textsubscript{2} in panel (a), showing suppression of the
electronic state at -0.4 eV (marked by an arrow). Set points: 2.0
V, 0.5 nA. Inset: $\mathrm{d}I/\mathrm{d}V$ spectrum recorded on
the bare Ir(111) surface, the state at -0.4 eV is present and is reminiscent
of the Ir surface state peak. Set points: -1.5 V, 0.5 nA. (c) $\mathrm{d}I/\mathrm{d}V$
map recorded at the energy of the -0.4 eV state. Green parallelograms
in (a) and (c) mark the unit cell of Bo. (d) $\mathrm{d}I/\mathrm{d}V$
line spectroscopy along the black line in the inset, showing modulation
of the electronic structure across consecutive Bo stripes. Set points:
2.0 V, 0.5 nA (-0.5 V, 0.5 nA for the inset).}

\end{figure*}

Further details of the electronic structure of Bo/Ir at the atomic
level are obtained from STM and STS measurement. A typical STM image
of the system is shown in Fig. \ref{figSTM-STS}(a), with a characteristic
stripe-like pattern of the Bo $\chi_{6}$ polymorph. The stripes of
width $d=14.0\pm0.1$ Å run diagonally across the figure panel and
appear to be separated by trenches. Additionally, each stripe consists
of two parallel rows of lobes, thus constituting two ``wavy'' sub-stripes
\citep{Vinogradov2019,Omambac2021,Cuxart2021}. For comparison, top-right
corner of Fig. \ref{figSTM-STS}(a) depicts the appearance of freestanding
Bo as calculated by DFT. In that case, the stripe-like pattern is
absent, which demonstrates that it is imprinted in the Bo/Ir system
by the Ir substrate. In order to investigate the contribution of the
electronic structure to such appearance of Bo on Ir(111), STS measurements
were carried out.

Fig. \ref{figSTM-STS}(b) shows two point spectra, recorded at the
lobe (P\textsubscript{1}) and trench (P\textsubscript{2}) positions.
Both spectra are V-shaped in the vicinity of $E_{\mathrm{F}}$ which
is a signature of metallicity of Bo/Ir. However, there is a notable
difference between the two spectra in the energy range between 0.3
and 0.5 eV below $E_{\mathrm{F}}$. The spectrum at the lobe position
contains a significantly higher density of states (DOS) in comparison
to the spectrum at the trench position {[}see arrow in Fig. \ref{figSTM-STS}(b){]}.
The same state was observed for the bare Ir(111) surface {[}see inset
of Fig. \ref{figSTM-STS}(b){]} and can be assigned to the surface
state Ir S\textsubscript{1} located at the $\Gamma$ point \citep{Schulz2014}.
Therefore, our data indicates that the surface state persists under
the Bo sheet, although with spatially-modulated intensity – the state
is suppressed in the trenches but is pronounced in the lobes of the
Bo’s stripe-like structure.

Such spatial modulation of the Ir surface state is even better visualized
in the STS map and line scan, shown in Fig. \ref{figSTM-STS}(c) and
(d). The STS map recorded at -0.4 V of bias voltage displays clear
coincidence with the topographic image from panel (a). This coincidence
is evidence of a link between Bo crystal structure and Ir surface
state coherence, which in turn points towards inhomogeneous binding
of Bo to Ir at the nanoscale. Such modulation of the Ir surface state
also explains its spectral modifications in our ARPES data.

\subsection{Electronic structure from DFT calculations}

\begin{figure*}
\begin{centering}
\includegraphics{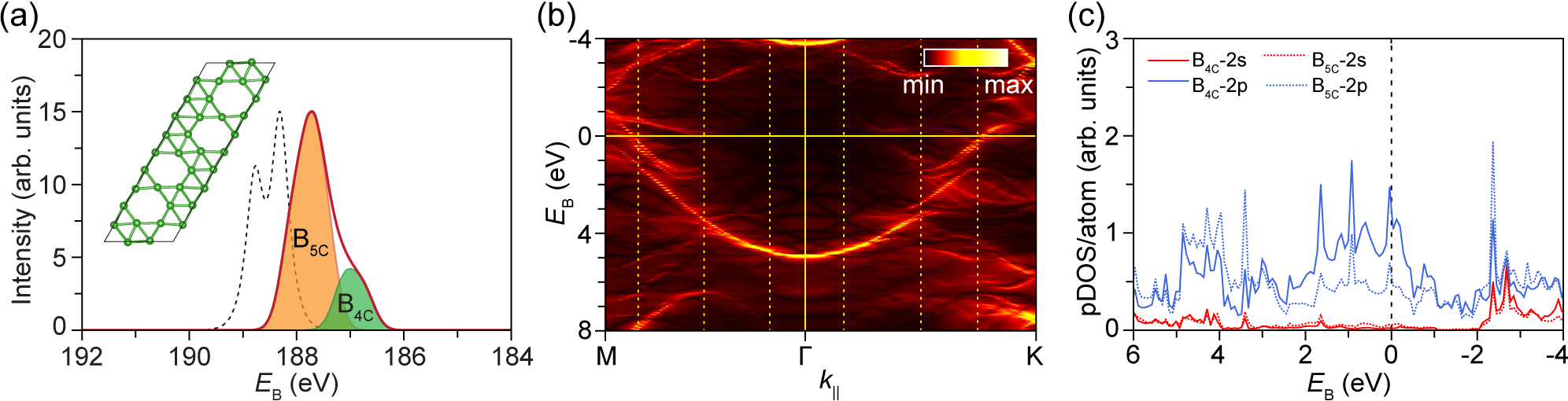}
\par\end{centering}
\caption{\label{fig:DFT}DFT calculations of electronic properties of a freestanding
Bo $\chi_{6}$ polymorph. (a) XPS spectra, with the two peaks arising
from 4- (B\protect\textsubscript{4C}) and 5-coordinated (B\protect\textsubscript{5C})
B atoms. Calculated $E_{\mathrm{B}}$ of each atomic site are combined
after being broadened by a Gaussian function with the standard deviation
of 0.3 eV. Dashed line is fitted curve to the experimental Bo/Ir data
form Fig. \ref{figXPS}(b)-bottom, added for comparison. Inset: structurally
optimized Bo unit cell. (b) Valence band structure of Bo averaged
over three possible Bo orientations and unfolded along M-$\Gamma$-K
high symmetry direction of the Ir SBZ. Vertical dashed lines mark
boundaries of the Bo SBZs. (c) Projected density of electronic states,
separated according to orbital character ($2s$ or $2p$) and coordination
number (4-coordinated 4C, or 5-coordinated 5C). All binding energies
are referenced to the Fermi level.}
\end{figure*}

DFT calculations of the electronic properties of freestanding $\chi_{6}$
Bo polymorph have been carried out. In this way we complement our
experimental data, in particular our ARPES and XPS findings, and provide
additional support to the nano-modulated nature of the Bo/Ir system.
Prior to the electronic structure calculation, the B atom positions
were relaxed in order to reach optimized crystal structure {[}see
Fig. \ref{fig:DFT}(a) inset{]}. The simulated XPS spectrum of the
B $1s$ peak of such an optimized structure is shown in Fig. \ref{fig:DFT}(a).
The spectrum essentially consists of two peaks, corresponding to five
4- (B\textsubscript{4C}) and twenty 5-coordinated (B\textsubscript{5C})
atoms. These two peaks are centered at 187 eV and 187.7 eV binding
energy, respectively. Altogether, the two peaks merge into one dominant
peak with a shoulder on the low binding energy side.

A fit to the experimental XPS data from Fig. \ref{figXPS}(b)-bottom
is also overlapped in Fig. \ref{fig:DFT}(a) as a dashed line. Apparently,
neither do the peak separation nor the peak intensity ratio match
the calculated expectations, which points towards nano-modulation
as a possible source of inhomogeneous Bo-Ir interaction. A shift of
the calculated spectrum towards lower binding energies with respect
to the experimental data is explained by the absence of the substrate
in our theoretical calculations.

The calculated band dispersion is plotted in Fig. \ref{fig:DFT}(b),
unfolded to the first SBZ of Ir(111) for a better overview. An average
of three 120°-rotated Bo domains is shown, thus mimicking area-averaging
of the conventional ARPES measurements (single-domain calculated bands
are shown in Supporting Figure S4). The most dominant feature is a
parabolic band with the vertex at $E_{\mathrm{B}}\sim5$ eV. Notably,
there are no pronounced electronic bands close to $E_{\mathrm{F}}$
around the $\Gamma$ point (and in particular, with a large velocity)
which could be identified as the Bo-induced bands from our ARPES measurements,
further confirming Ir band replication scenario. The corresponding
projected density of states (pDOS) is plotted in Fig. \ref{fig:DFT}(c).
It is clear that Bo bands are dominantly of $p-$character, for both
4- and 5-coordinated B atoms. Published calculations of other Bo polymorphs
($\chi_{3}$ and $\beta_{12}$) show similarity to these results,
most notably in respect to the orbital symmetry and binding energy
of the $p-$character bands centered at the $\Gamma$ point \citep{Penev2016,Abbasi2023}.
The calculated work function of a freestanding Bo layer is $\Phi_{\mathrm{Bo,free}}$
= 4.86 eV.

\section{Discussion}

Let us first clarify the origin of the B\textsubscript{1} and B\textsubscript{2}
peaks found in the XPS spectra of a fresh Bo/Ir sample. Considering
all of our data, we conclude that the two B $1s$ peaks arise from
distinct registries and inhomogeneous binding of B atoms with respect
to the substrate atoms, similar as for hBN on Ir(111) and on other
metallic substrates as well \citep{Orlando2011,Preobrajenski2007}.
In our case, B atoms located in the lobes and trenches exhibit notably
different interaction with Ir, in agreement with the STS data in Fig.
\ref{figSTM-STS}, which is then reflected in two distinct B $1s$
levels. In analogy with the hBN/Ir(111) system \citep{Orlando2011},
it is plausible that the lower binding energy component B\textsubscript{1}
corresponds to B atoms located in the weakly bound lobes. Those B
atoms readily react with residual molecules from UHV {[}cf. Fig. \ref{figXPS}(b){]}.
The higher binding energy component B\textsubscript{2} originates
from B atoms strongly bound to Ir within the trenches of the Bo structure
and they are somewhat more robust against ``rapid'' adsorption of
molecules from the UHV. Similar assignment of the two B $1s$ peaks,
separated by $\sim0.8$ eV, was assumed before for buckled Bo on Ag(111)
\citep{Mannix2015}. Therefore, B-Ir interaction exhibits nano-modulation
and in parallel induces spatially-varying chemical reactivity of Bo
from the top (i.e., vacuum) side. In addition, inhomogeneous B-Ir
interaction induces the stripy structure of Bo, as also supported
by the simulated STM image of a freestanding $\chi_{6}$ Bo {[}inset
in Fig. \ref{figSTM-STS}(a){]}.

Other factors, such as coordination of B atoms with other B neighbors,
cannot be the reason for the double-peaked B $1s$ structure primarily
since this effect should yield $I_{\mathrm{B1}}/I_{\mathrm{B2}}=4$,
in line with our DFT calculations in Fig. \ref{fig:DFT}(a), which
is far from the measured experimental ratio. Even though the intrinsic
coordination effect may exist, we presume that it is obscured in our
data by the establishment of B-Ir bonds and dominance of the nonuniform
Bo-Ir interaction described above. Also, we eliminate B atoms still
embedded in Ir (sub)surface as potential origin of the B $1s$ double-peak
structure, as in the case of Bo on Al(111) \citep{Preobrajenski2021}.
This scenario is unlikely since both B\textsubscript{1} and B\textsubscript{2}
disappear upon Bo oxidation in ambient conditions \citep{Jugovac2023},
whereas at least one component (corresponding to the alloy) should
remain protected. Also, we found no evidence of a new Ir $4f$ peak
which would signify the presence of B-Ir alloy, as in the case of
favorable Si-Ir intermixing over silicene formation \citep{Satta2018}.
Contributions from other Bo polymorphs on the sample surface to the
double-peak B $1s$ structure are excluded based on our STM, LEEM
and LEED data.

Besides an evident modulation of the B-Ir interaction at the nanoscale,
we can also qualitatively assess B-Ir binding from our XPS, ARPES
and STS data. The reduction of the iridium core-level component Ir\textsubscript{S}
{[}Fig. \ref{figXPS}(a){]} and the suppression of the Ir surface
state Ir S\textsubscript{1} {[}Fig. \ref{figCARPES}, Fig. \ref{figUARPES}
and Supporting Figure S3{]} signify a rather strong B-Ir interaction.
However, the state Ir S\textsubscript{1} is not completely quenched
under the Bo layer (Fig. \ref{figSTM-STS}), which would altogether
categorize the B-Ir interaction as moderate. Moderate interaction
is in line with the ability of room temperature intercalation of gold
atoms underneath Bo on Ir(111) \citep{Vinogradov2019}. Interestingly,
the surface state of Ag was quenched under the Bo layer \citep{Liu2019},
which might indicate stronger binding of Bo to Ag(111) as compared
to Ir(111).

Investigation of the valence band structure of Bo is hindered by substantial
chemical (re)activity of Bo even in UHV conditions (cf. Fig. \ref{figXPS}).
Additionally, the rich electronic structure of Ir obscures (and potentially
hybridizes with) any band which does not fall into one of Ir's bandgaps
\citep{Pletikosic2010}. In this study, clearly visible new features
arising from the Bo presence on the sample surface are Ir replica
bands. Opening of electronic gaps in their vicinity has not been observed
(as opposed to graphene on Ir(111) \citep{Pletikosic2009}), which
confirms their umklapp origin and also final-state-effect character.
Considering the common sources of such band replication, one can assume
that the reciprocal lattice of Bo is the source of electron scattering.
It can be immediately recognized that the replication wavevector $\mathbf{q}$
matches (both in direction and magnitude) the reciprocal vectors $3\mathbf{b}_{1}$
or $\mathbf{b}_{2}$ of the three Bo domains {[}see Fig. \ref{figCARPES}(c){]},
since $3b_{1}=b_{2}=\pi/\left[a_{\mathrm{Ir}}\cos\left(30{^\circ}\right)\right]=1.34$
Å\textsuperscript{-1} where $a_{\mathrm{Ir}}=2.715$ Å is the surface
lattice constant of Ir(111) \citep{Arblaster2010}. However, if either
$3\mathbf{b}_{1}$ or $\mathbf{b}_{2}$ vectors alone would cause
the umklapp scattering, the number of the replicated bands in ARPES
data should be rather large, as depicted in Fig. \ref{figCARPES}(e)
(see Supporting Figure S5 for additional details). Since this situation
does not correspond to our ARPES measurements, some other factors
must be involved in the electron scattering process.

A reasonable assumption is that the stripe-like structure of Bo can
affect electron scattering, since it introduces an additional periodicity
in the system. If only scattering of electrons with momentum perpendicular
to the Bo stripes is taken into account, the number of replicas is
significantly reduced. The expected replica arrangement in that case
is schematically shown in Fig. \ref{figCARPES}(d), which is in agreement
with our experimental measurements. Therefore, umklapp process is
not induced by the wavevectors $3\mathbf{b}_{1}$ or $\mathbf{b}_{2}$,
but rather by the stripy structure of Bo which constitutes a 1D lattice
with a wavevector $\mathbf{q}$. Scattering is happening only in the
direction perpendicular to the Bo stripes, i.e. only for the electrons
moving along $\left\langle 11\overline{2}\right\rangle $ directions.
Similar 1D-only umklapp processes have been identified previously
for graphene nanoribbons on Ni(771) \citep{Shikin2003}, and graphene
sheets on Pt(997) \citep{Pisarra2014} and reconstructed Au(001) \citep{Terasawa2023}.
To the best of our knowledge, band replication induced by atomically
thin boron layers has not been reported before. Despite the electron
scattering on Bo stripe structure, the relation $\mathbf{q}=3\mathbf{b}_{1}$
still holds, meaning that the corresponding diffraction spots in the
LEED data (stripe-related and lattice-related) cannot be distinguished.

The replication vector $\mathbf{q}$ translates into a real-space
periodicity of $2\pi/q=4.69\pm0.03$ Å. This value coincides with
$d/3$, i.e. one third of the Bo stripe width. In the STM and STS
data in Fig. \ref{figSTM-STS}, this periodicity can be recognized
as a characteristic lengthscale of Ir topography and surface state
modulation in the direction perpendicular to Bo stripes, thus being
closely related to the sub-stripes. Therefore, unidirectional nano-scale
modulation of the Bo-Ir binding perturbs Bo lattice in a way to yield
a 1D stripy structure – i.e. an atomically-thin diffraction grating
– with a wavevector $\mathbf{q}$ as the dominant scattering component.

\section{Conclusions}

In a combined experimental and theoretical approach, we investigated
the structural and electronic properties of an extended borophene
(Bo) monolayer on the (111) surface of Ir. Significant nanoscopic
and unidirectional modulation of Bo is found, arising form nonuniform
interaction of B atoms with the underlying Ir atoms. Such modulation
results in a stripe-like structure of Bo, which in turn acts as an
electron diffraction grating. Electron diffraction (i.e. scattering)
is manifested in the form of replica bands which have been detected
in the photoemission spectra of the Bo/Ir system in the vicinity of
the Fermi level. The pristine structure of this atomically-thin boron
grating, being chemically (re)active and prone to atomic-scale modifications,
can be easily restored by executing a dissolution-segregation cycle
during which B atoms reassemble into a fresh Bo mesh.

Our results showcase scattering capabilities of 2D materials and stimulate
further examination of other Bo polymorphs which might exhibit different
types of superperiodicity phenomena. The presented data constitute
a comprehensive insight into the electronic structure of epitaxial
Bo which is vital for further development of Bo-based applications.
In addition, our work demonstrates facile fabrication of a nanoscopic,
highly-ordered system which may serve as a template for further functionalization
and also become a building block in more complex 2D-based heterostructures
and devices with enhanced electronic properties.

\section{Methods}

\subsection{Sample preparation}

Bo samples were synthesized in four ultra-high vacuum (UHV) systems
on an Ir(111) single-crystal (see below for details of these systems),
and sample preparation procedure was the same in all of them, with
minor differences which yielded the same end result. Ir substrate
cleaning consisted of argon or xenon sputtering at $1.5-2.5$ keV
followed by heating in oxygen atmosphere at $800-1000$ °C and annealing
at $1100-1250$ °C. The surface quality was checked by sharp low-energy
electron diffraction (LEED) spots of Ir, presence of the Ir surface
states as well as absence of contamination-related peaks in XPS (where
applicable). For Bo synthesis, boron dissolution-segregation method
was used \citep{Omambac2021,Radatovic2022}, during which the Ir surface
was repeatedly exposed to borazine at a pressure in the range $1\times10^{-8}-5\times10^{-7}$
mbar and temperature of $990-1100$ °C, and subsequently cooled to
room temperature. Sample heating was realized via e-beam bombardment
(with typical heating rates in the 15-20 °C/s range) while cooling
was achieved by switching off the e-beam source (with typical cooling
rates in the 3-10 °C/s range), without any additional active cooling
mechanism. Sample preparation parameters used in different setups
are given in the Supporting Table 1.

\subsection{X-ray photoelectron spectroscopy (XPS) }

XPS data were recorded \textit{in situ} at the VUV-Photoemission beamline
of the Elettra synchrotron (Trieste, Italy) at room temperature and
in normal emission. A Scienta R4000 electron analyzer was used for
data acquisition with the slit oriented parallel to the scattering
plane of the experiment. Linearly polarized light illuminated the
sample at 45° with respect to the analyzer axis, with the light spot
diameter on the sample surface of $\sim0.5$ mm. XPS peaks were fitted
by combined Gaussian-Lorentzian functions and Shirley background.
The binding energy scale was calibrated either by using Ir $4f$ bulk
peaks or the Fermi level.

\subsection{Conventional angle-resolved photoemission spectroscopy (ARPES)}

ARPES data were recorded \textit{in situ} in two setups: the VUV-Photoemission
beamline of the Elettra synchrotron (Trieste, Italy) and the UARPES
(now URANOS) beamline of the Solaris synchrotron (Krakow, Poland).
At VUV-Photoemission, data {[}Fig. \ref{figCARPES}, Supporting Figure
S2(d-f) and Supporting Figure S3(a){]} were recorded at the same conditions
and geometry as described in the preceding (XPS) subsection. 4D data
sets ($k_{x},k_{y},E,I$) were acquired by rotating the sample azimuthally
and by applying symmetry operations to the data. At UARPES, data {[}Supporting
Figure S2(a-c){]} were recorded \textit{in situ} and at room temperature
by employing the Scienta DA30L electron analyzer. Linearly polarized
light illuminated the sample at 44° with respect to the analyzer axis,
with the light spot diameter on the surface of $\sim0.5$ mm. 4D data
sets ($k_{x},k_{y},E,I$) were acquired by exploiting the full acceptance
cone of the analyzer, i.e. no sample movement was necessary.

\subsection{Micro-spot angle-resolved photoemission spectroscopy ($\mu-$ARPES)}

$\mu-$ARPES measurements were performed \textit{in situ} at the Nanospectroscopy
beamline of the Elettra synchrotron (Trieste, Italy) with the SPELEEM
III microscope (Elmitec GmbH) equipped with a hemispherical analyzer
(Elmitec R200) and a 2D detector comprising a phosphorous screen and
a CCD camera (QImaging Retiga). The sample was either illuminated
with low-energy electrons (LEEM and LEED mode) or photons (ARPES mode).
By inserting apertures in the illumination or imaging column of the
microscope, characterization of the crystallography and electronic
structure on the micrometer scale becomes available ($\mu-$LEED and
$\mu-$ARPES, probing area of $\sim2$ $\mu$m in diameter). The electron
energy is set by applying a voltage bias to the sample, commonly referred
to as start voltage ($V_{\mathrm{st}}$).

\subsection{Scanning tunneling microscopy and spectroscopy (STM and STS)}

STM/STS measurements were carried out \textit{in situ} at room temperature
in a home-built system ``Athene'' at the University of Cologne,
Germany. The $dI/dV$ curves were acquired by numerically differentiating
single-point $I/V$ spectra, captured in 128\texttimes 128 point grids.
The processing of STM images (background subtraction, contrast enhancement)
was done with the WSxM software \citep{Horcas2007}.

\subsection{Density functional theory (DFT) calculations}

We performed first-principles calculations based on DFT as implemented
in the OpenMX code \citep{Ozaki2003}. The exchange-correlation effect
in electrons was addressed by the generalized gradient approximation
\citep{Perdew1996}. Each B atom had a $s3p2d1$ pseudoatomic-orbital
basis set with 7.0 bohr cutoff radius. By setting the bulk Ir lattice
constant to 3.877 Å, we utilized a slab model for the free-standing
borophene, whose in-plane lattice parameter was fixed to the Ir(111)$-\left(6\times2\right)$
lattice. The distance between slabs was 12 Å. In the geometry optimization,
$10^{-6}$ hartree and $10^{-5}$ hartree/bohr of the total-energy
and force-convergence criteria were utilized, respectively. The $\Gamma$-centered
$6\times3\times1$ $k-$grid was adopted. The core-hole pseudopotential
on B $1s$, with the addition of a penalty-functional and Coulomb-cutoff
approach as proposed by Ozaki et al. \citep{Ozaki2017}, was used
for the calculations of the core-level excitation. To minimize the
artificial interaction between core holes, the supercell corresponding
to the Ir(111)$-\left(6\times6\right)$ lattice was incorporated into
calculations. Within the Tersoff-Hamann approach, simulations of the
STM images were carried out for an energy window between the Fermi
level and -0.5 eV from it \citep{Tersoff1985,Horcas2007}. For the
band-structure unfoldings to the Ir(111) first SBZ \citep{Wang2021},
the projector-augmented wave method was utilized as implemented in
the VASP code \citep{Blochl1994,Kresse1996,Kresse1999}.

\subsection*{Supporting information}

The supporting information is available free of charge at https://pubs.acs.org/doi/10.1021/acsami.3c14884.
LEEM data sequence of borophene dissolution and recondensation, additional
ARPES data of clean Ir and Bo/Ir samples, energy distribution cuts
(EDCs) at the $\Gamma$ point before and after Bo formation, band
dispersions of freestanding Bo in different high-symmetry directions
as calculated by DFT, detailed schematics of the Fermi surface for
different scattering scenarios, table of experimental parameters used
for sample preparation.

\subsection*{Author contributions}

S.K., P.M., P.M.S., A.V.M., M.K., and M.P. performed ARPES and XPS
experiments. S.K., I.Š.J. and M.P. analyzed the ARPES and XPS data.
M.J., C.A.B., T.O.M. and A.L. conducted $\mu-$ARPES and LEEM measurements
and analyzed the respective data. I.S. and Y.G. executed and interpreted
DFT calculations. P.B. and T.M. conducted and interpreted STM/STS
experiments. All co-authors discussed the results. M.P. and S.K. prepared
the manuscript with contributions from all co-authors. M.P. conceived
and supervised research. All authors have given approval to the final
version of the manuscript.
\begin{acknowledgments}
This work was supported by the Croatian Science Foundation, Grant
No. UIP-2020-02-1732 and by the Center of Excellence for Advanced
Materials and Sensing Devices, ERDF Grant No. KK.01.1.1.01.0001. P.B.
gratefully acknowledges financial support from the Alexander von Humboldt
Foundation. T.M. acknowledges support by DFG within the project “Cluster
Superlattice Membranes” (project No. 452340798). We acknowledge Elettra
Sincrotrone Trieste for providing access to its synchrotron radiation
facilities and for financial support under the IUS (P2022003) project.
P.M., P.M.S. and A.V.M. acknowledge EUROFEL-ROADMAP ESFRI of the Italian
Ministry of Education, University, and Research. This publication
was developed under the provision of the Polish Ministry of Education
and Science project: ``Support for research and development with
the use of research infrastructure of the National Synchrotron Radiation
Centre SOLARIS'' under contract No. 1/SOL/2021/2. We acknowledge
SOLARIS Centre and in particular staff of the UARPES beamline for
the access to the beamline and assistance during the measurements.
The authors acknowledge the CERIC-ERIC Consortium for the access to
experimental facilities and financial support. The calculations were
partly carried out by using supercomputers at ISSP, The University
of Tokyo, and TSUBAME, Tokyo Institute of Technology.
\end{acknowledgments}

\bibliographystyle{apsrev4-1}
\bibliography{references}

\end{document}